\documentclass[amstex,epsf,amssymb,usenatbib]{mn2e}
\usepackage{epsf,graphics}
\usepackage{amsmath,amssymb,natbib}
\usepackage{rotating,times,pictex,graphicx,latexsym,color,longtable}
\def\aj{AJ}                   % Astronomical Journal
\def\apj{ApJ}                 % Astrophysical Journal
\def\aap{A\&A}                % Astronomy and Astrophysics
\def\balta{BaltA}             % Baltic Astronomy
\def\dossr{DoSSR}             % Akademiia Nauk SSSR Doklady
\def\mnras{MNRAS}             % Monthly Notices of the RAS
\def\pasj{PASJ}               % Publications of the ASJ
\def\phrvl{PhRvL}             % Physical Review Letters
\def\hsf{HSFR~VI}             % Handbook of Star Forming Regions, Volume I

\title{The Structure of Molecular Clouds: II - Column Density and Mass
Distributions}

\author[D. Froebrich, J. Rowles]  {Dirk Froebrich $^{1}$\thanks{E-mail:
df@star.kent.ac.uk, } Jonathan Rowles $^{1}$\thanks{E-mail: jr262@kent.ac.uk}\\
$^1$Centre for Astrophysics \& Planetary Science, The University of Kent,
Canterbury, Kent CT2 7NH, U.K.} 

\date{Accepted .....
      Received ..... ;
      in original form .....}

\pagerange{\pageref{firstpage}--\pageref{lastpage}}
\pubyear{2010}

\begin{document}
\maketitle
\label{firstpage}

\begin{abstract} 

The formation of stars is inextricably linked to the structure of their 
parental molecular clouds. Here we take a number of nearby giant molecular
clouds (GMCs) and analyse their column density and mass distributions. This
investigation is based on four new all-sky median colour excess extinction maps
determined from 2MASS. The four maps span a range of spatial resolution of a
factor of eight. This allows us to determine cloud properties at a common 
spatial scale of 0.1\,pc, as well as to study the scale dependence of the cloud
properties.

We find that the low column density and turbulence dominated part of the clouds
can be well fit by a log-normal distribution. However, above a universal 
extinction threshold of $6.0 \pm 1.5$\,mag $A_V$ there is excess material 
compared to the log-normal distribution in all investigated clouds. This 
material represents the part of the cloud that is currently involved in star
formation, and thus dominated by gravity. Its contribution to the total mass of
the clouds ranges over two orders of magnitude from 0.1 to 10\,\%. This implies
that our clouds sample various stages in the evolution of GMCs. Furthermore, we
find that the column density and mass distributions are extremely similar
between clouds if we analyse only the high extinction material. On the other
hand, there are significant differences between the distributions if only the
low extinction, turbulence dominated regions are considered. This shows that the
turbulent properties differ between clouds depending on their environment.
However, no significant influence on the predominant mode of star formation
(clustered or isolated) could be found. Furthermore, the fraction of the cloud
actively involved in star formation is only governed by gravity, with the column
density and mass distributions not significantly altered by local feedback
processes.

\end{abstract}

\section{Introduction}
\label{intro}

The study of Giant Molecular Clouds (GMCs) is of great importance to understand
star formation. Their properties (density distribution, dynamics, temperature,
etc.) are thought to be vital in determining whether, where and how stars are
forming within them. Turbulent motions inside a cloud cause fragmentation and
hence determine the density distribution (Padoan et al.
\cite{1997ApJ...474..730P}, Ballesteros-Paredes et al.
\cite{1999ApJ...527..285B}). This in turn should have an influence on which mode
of star formation (isolated or clustered) is occurring (Klessen, Heitch and Mac
Low \cite{2000ApJ...535..887K}). It is, however, unclear if this is the sole
determinant of the star formation mode. Or are there other important causes such
as the environment, feedback and/or magnetic fields? 

One way to try to answer these questions is to investigate the column  density
structure of a large number of GMCs in a systematic, comparable and unbiased
fashion. This means ideally one should use the same tracer of column density for
all clouds. Such a tracer should be as bias free as possible and all clouds
should be investigated at the same physical spatial resolution. There are a
number of techniques to determine the column density of clouds. These include
molecular line emission, dust continuum emission, scattered infrared light or
extinction measurements. The latter on the bases of star counts, colour excess
or combined methods. See Froebrich \& Rowles \cite{2010BASI..........F} for a
brief summary and discussion of the advantages and disadvantages of the various
techniques. It has been shown by Goodman et al. \cite{2009ApJ...692...91G} that
near infrared extinction mapping is the best tracer of the `real' column density
of a cloud, based only on the assumption of a constant gas to dust ratio.

A number of large scale maps of the column density distribution of clouds have
been made to date, which could be used to compare the structure of a number of
GMCs. These are for example the maps by Cambr\'{e}sy \cite{1999A&A...345..965C},
Schlegel et al. \cite{1998ApJ...500..525S}, Dobashi et al.
\cite{2005PASJ...57S...1D}, Froebrich et al. \cite{2007MNRAS.378.1447F} and
Rowles and Froebrich \cite{2009MNRAS.395.1640R}. Some of them do not use the
ideal tracer (extinction) for the column density. Others use star counts, which
result in a distance dependent bias in the column density distribution
(Froebrich \& del Burgo \cite{2006MNRAS.369.1901F}), similar to using the mean
colour excess instead of the median. One bias they all have in common is that
clouds at different distances are mapped at different spatial resolutions. This
will naturally lead to a change in the observed column density distribution,
since for more distant clouds one averages over larger physical scales. This
will lead to the effect that small scale high extinction regions are averaged
out and the column density distributions are skewed. In order to compare the
structure of GMCs with each other, one has to determine the column density
distribution at similar physical scales and using the same `good' tracer for all
investigated clouds.

Only then can we attempt to draw conclusions about similarities and differences
in their structures. This can be done by analysing the column density and mass
distribution (the focus of this paper) or determining structure functions (the
focus of the next paper in the series) and comparing it to model predictions
such as published by  Kolmogorov \citep{1941DoSSR..30..301K}, She \&  Leveque
\citep{1994PhRvL..72..336S} and Boldyrev \citep{2002ApJ...569..841B}. In our
previous paper (Rowles \& Froebrich \cite{2009MNRAS.395.1640R}, Paper\,I
hereafter) we presented all-sky  extinction maps determined using data from
2MASS (Skrutskie et al. \cite{2006AJ....131.1163S}). These were created using an
adaptation of the NICE method (Lada et al. \cite{1994ApJ...429..694L}), median
colour excess determination and variable spatial resolution to obtain a constant
signal to noise ratio. Here we present and analyse additional maps, determined
using a range of constant spatial resolutions to map different clouds at similar
spatial scales. 

In Sect.\,\ref{method} we briefly describe the new maps created for the project,
and explain the data analysis methods. In Sect.\,\ref{results} we present the
results for our selection of nearby GMCs. We discuss these results and draw
conclusions in Sect.\,\ref{dissc}. 

\begin{table}
\centering
\caption{\label{cloudlst} Summary table of the clouds analysed in this paper. We
list the cloud name, the range in galactic coordinates ($l$;$b$) covered by the
region and the distance (d) adopted in this paper. $^*$The cloud Auriga\,1 is
referred to as the California Molecular Cloud in Lada et al. (2009). The
references for the adopted distances are the following: $^1$ Lada et al. (2009);
$^2$ Kenyon et al. (2008); $^3$ Kun (2008); $^4$ Knude \& Hog (1998); $^5$
Bally et al. (1999); $^6$ Muench (2008); $^7$ Lombardi et al. (2008); $^8$
Carpenter (2008); $^9$ Motte (2001); $^{10}$ Straizys (1996).}
\begin{tabular}{llll}
Name & $l$ range & $b$ range & d [pc] \\ \hline
Auriga\,1$^*$    &      156$\degr$ --           173$\degr$ &      $-$12.0$\degr$ -- \,\,\,$-$3.0$\degr$ & 450$\pm$\,\,\,23$^1$     \\
Auriga\,2        &      175$\degr$ --           186$\degr$ &      $-$11.0$\degr$ -- \,\,\,$-$2.0$\degr$ & 140$\pm$\,\,\,28$^2$     \\
Cepheus          &      100$\degr$ --           120$\degr$ & \,\,\,$+$5.0$\degr$ --      $+$20.0$\degr$ & 390$\pm$125$^3$          \\
Chamaeleon       &      296$\degr$ --           304$\degr$ &      $-$18.0$\degr$ --      $-$13.0$\degr$ & 150$\pm$\,\,\,30$^4$     \\
Circinus         &      316$\degr$ --           319$\degr$ & \,\,\,$-$6.0$\degr$ -- \,\,\,$-$3.0$\degr$ & 700$\pm$350$^5$          \\
Corona Australis &      359$\degr$ -- \,\,\,\,\,\,1$\degr$ &      $-$20.0$\degr$ --      $-$17.0$\degr$ & 170$\pm$\,\,\,34$^4$     \\
$\lambda$-Ori    &      188$\degr$ --           201$\degr$ &      $-$18.0$\degr$ -- \,\,\,$-$6.0$\degr$ & 400$\pm$\,\,\,80$^6$     \\
Lupus\,1, 2      &      334$\degr$ --           342$\degr$ &      $+$11.0$\degr$ --      $+$19.0$\degr$ & 155$\pm$\,\,\,\,\,\,8$^7$\\
Lupus\,3, 4, 5, 6&      335$\degr$ --           344$\degr$ & \,\,\,$+$5.0$\degr$ --      $+$11.0$\degr$ & 155$\pm$\,\,\,\,\,\,8$^7$\\
Monoceros        &      212$\degr$ --           222$\degr$ &      $-$13.0$\degr$ -- \,\,\,$-$7.0$\degr$ & 830$\pm$\,\,\,50$^8$     \\
Ophiuchus        &      350$\degr$ --           360$\degr$ &      $+$12.0$\degr$ --      $+$19.0$\degr$ & 119$\pm$\,\,\,\,\,\,6$^7$\\
Orion\,A         &      208$\degr$ --           219$\degr$ &      $-$21.0$\degr$ --      $-$16.0$\degr$ & 410$\pm$\,\,\,80$^6$     \\
Orion\,B         &      201$\degr$ --           211$\degr$ &      $-$17.0$\degr$ -- \,\,\,$-$8.0$\degr$ & 410$\pm$\,\,\,80$^6$     \\
Perseus          &      156$\degr$ --           163$\degr$ &      $-$25.0$\degr$ --      $-$15.0$\degr$ & 310$\pm$\,\,\,65$^9$     \\
Serpens          & \,\,\,30$\degr$ --      \,\,\,32$\degr$ & \,\,\,$+$4.0$\degr$ -- \,\,\,$+$6.0$\degr$ & 260$\pm$\,\,\,10$^{10}$  \\
Taurus           &      164$\degr$ --           178$\degr$ &      $-$19.0$\degr$ --      $-$10.0$\degr$ & 140$\pm$\,\,\,28$^2$     \\ 
\end{tabular}
\end{table}

\section{Data Analysis}
\label{method}

\subsection{Cloud selection}
\label{cloudsel}

For our analyses we selected a number of nearby GMCs. Since we are interested in
their column density distribution, we selected only clouds that are not situated
directly in the Galactic Plane. This avoids that there are two clouds along the
same line of sight, rendering the column density analysis difficult.
Furthermore, only sufficiently large (or nearby) clouds are selected. This
ensures that we have enough area (pixels) in our maps to analyse the column
density distribution. Finally, only GMCs with a well known distance are included
in our investigations. This selection process leaves 16 cloud complexes. Their
names, coordinate ranges, distances and references are listed in
Table\,\ref{cloudlst}. To allow a comparison to the earlier work by Froebrich et
al. \cite{2007MNRAS.378.1447F} we adopt the names for the regions from this
paper. In the case of the Lupus complex, we treat several of the small clouds as
one, in order to have enough area to be analysed. One should hence keep in mind
when interpreting this data that they are composed of several clouds.

\subsection{Constant resolution extinction maps}
\label{fixmaps}

The determination of the new constant resolution extinction maps has been
performed in exactly the same way as described in Paper\,I. We determine median
$\left< J-H \right>$ and $\left< H-K \right>$ colour excess maps, with the same
position dependent zero point as in Paper\,I. These maps are converted into
optical extinction maps and are averaged (following Eq.\,\ref{eqconversion} and
using $\beta$\,=\,1.7 as in Paper\,I). Similarly, maps of the uncertainties are
calculated. Already in Paper\,I one constant resolution map has been determined.
There the spatial resolution changed from 0.5\arcmin\ to 2.0\arcmin, depending
on the galactic latitude. Only stars within this radius were included in the
extinction determination. In contrast to this, the new maps consist of square
shaped pixels with no oversampling. In other words, the pixel values in those
maps are completely independent on the neighbouring pixels.

\begin{equation}\label{eqconversion}
A_V = \frac{5.689}{2} \cdot \left( \frac{\left< J-H \right>}{\left(
\frac{\lambda_H}{\lambda_J} \right)^\beta - 1} + \frac{\left< H-K \right>}{1 -
\left( \frac{\lambda_K}{\lambda_H} \right)^{-\beta}} \right)
\end{equation}

For the purpose of this paper, we have calculated a number of further constant
spatial resolution maps, in total four, with a different resolution. We used the
resolution of the maps in Paper\,I for the first map, and then increased the
pixel size in steps of a factor of two. This ensures we have an extinction map
of each cloud with spatial resolutions covering almost an order of magnitude.
Since the range of distances for our clouds varies by less than that, there will
be one common physical spatial resolution where each cloud has been observed. 

The new all-sky extinction maps available are listed in Table\,\ref{avail3}.
They are labeled maps\,1 through 4 depending on the pixel sizes used. Note the
dependence of the actual pixel size of each map with galactic latitude. 

\begin{table}
\caption{\label{avail3} Pixel sizes of the new all-sky constant resolution
extinction maps.}
\centering
\begin{tabular}{l|cccc}
$|b|$ range & Pixel size & Pixel size & Pixel size & Pixel size \\
 & Map\,1\,[\arcmin] & Map\,2\,[\arcmin] & Map\,3\,[\arcmin] & Map\,4\,[\arcmin] \\ \hline
90$\degr$ -- 50$\degr$      & 2.0 & 4.0 & 8.0 &    12.0 \\
50$\degr$ -- 40$\degr$      & 1.5 & 3.0 & 6.0 &    12.0 \\
40$\degr$ -- 20$\degr$      & 1.0 & 2.0 & 4.0 & \,\,8.0 \\
20$\degr$ -- \,\,\,0$\degr$ & 0.5 & 1.0 & 2.0 & \,\,4.0 \\ \hline
\end{tabular}
\end{table}

\subsection{Cloud structure analysis}
\label{colden}

We can now determine the column density distribution for each cloud at four
different spatial scales. As a first step we extract the extinction values for
each cloud from the maps\,1 to 4. We then plot histograms of the number of
pixels with a given extinction value, or the integrated number of pixels above a
given extinction -- i.e. a measure of the mass distribution.

The width of the histograms bins was varied. We used 1/8th, 1/4th, 1/2 and
1\,mag of optical extinction. All subsequent analyses were conducted for each
bin size, in order to check for systematic dependence of the results on this. It
turned out that, as long as there are a sufficient number of pixels in each bin,
the results do not depend on the bin size. For small clouds and large spatial
resolution the last bin width (1\,mag) might contain not enough datapoints. We
hence excluded these outliers and averaged the results obtained from the other
histogram bin widths to calculate the final result.

\subsubsection{Log-Normal Fits to the Column Density Distribution}
\label{fittwo}

The first method of analysis performed was to fit an analytic function to the
column density distribution histogram showing the number of pixels $N$ with an
optical extinction $A_V$. Following Lombardi et al. \cite{2008A&A...489..143L}
we fit our clouds with a log-normal distribution of the form:

\begin{equation}
\label{lom12}
h(A_V) = \frac{a}{{A_V - A_0}} \exp \left[ - \frac{ \left( \ln  \left({A_V -
A_0} \right) - \ln{A_1} \right)^2}{2\left(\ln{\sigma}  \right)^2} \right]
\end{equation}

If there are only a small number of clouds along the line of sight (certainly
valid due to our selection criteria of clouds; see Sect.\,\ref{cloudsel}) and
the underlying density distribution is log-normal (as predicted e.g. by
V\'{a}zquez-Semadeni \& Garcia \cite{2001ApJ...557..727V}) a good fit should be
obtained.

\begin{figure*} 
\centering
\includegraphics[width=8cm]{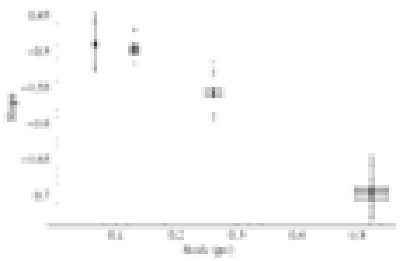}\hfill
\includegraphics[width=8cm]{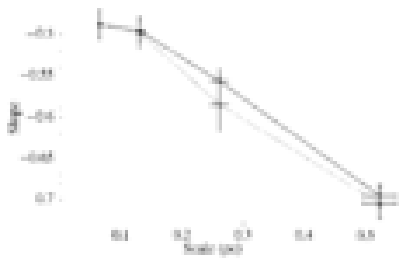}
\caption{\label{Aur1FS} {\bf Left:} A plot showing measured slopes $\gamma_{\rm
low}$ against spatial scale for the Auriga\,1 cloud. The four data points at
each resolution correspond to the four histogram bin widths used. {\bf Right:}
The solid line shows the averaged slopes $\gamma_{\rm low}$ from the left panel.
The dotted line shows the slopes $\gamma_{\rm low}$ if the highest resolution
image is simply rebinned to the lower resolutions.}
\end{figure*}

\subsubsection{The $\log(N)$ vs $A_V$ Column Density Distribution} 
\label{fitone}

A plot of the optical extinction vs $\log(N)$, where $N$ is the number of pixels
with the given $A_V$ value, is to a large extent linear. This is expected since
the column density distribution is caused by turbulent fragmentation and
turbulence is intrinsically self-similar. We can hence use the slope ($\gamma$)
of this distribution to characterise the column density distribution.

For almost all clouds there are two ranges of extinction where the plot is
linear. Both possess different slopes $\gamma$ and typically the change of slope
happens between 5 and 10\,mag of $A_V$. We hence determine for each cloud two
slope values, one for the lower extinction region ($\gamma_{\rm low}$) and one
for the high $A_V$ material ($\gamma_{\rm high}$). The extinction ranges where
the slopes are constant are set for each cloud manually.

\subsubsection{The $\log(M)$ vs $A_V$ Mass Distribution}
\label{masscalc}

Similar to the analysis of the column density distribution we can calculate the
mass $M$ in the cloud at extinction values higher than a given $A_V$ value.  In
other words we integrate the extinction values above $A_V$ and convert to masses
using the pixel size and distance to the cloud. Like for the investigation of
the column density, this mass distribution shows, over a range of $A_V$ values,
an exponential behavior. We can hence fit the slope ($\delta$) in the $\log(M)$
vs $A_V$ diagram to characterise the mass distribution. 

Again, for most clouds there are two regimes with different slopes. There is the
lower $A_V$ region (characterised by $\delta_{\rm low}$), whose mass
distribution is governed by the turbulent properties of the cloud material. And
the high $A_V$ region (characterised by $\delta_{\rm high}$), where gravity
becomes important and changes the mass distribution. The extinction value where
this change of behavior is observed (A$_{V,SF}$), can be seen as the extinction
threshold for star formation, first described by Johnstone et al.
\cite{2004ApJ...611L..45J} in Ophiuchus. 

\section{Results}
\label{results}

\subsection{Scale Dependent Effects}

We are planning to compare parameters of cloud structure for different clouds.
Hence, all parameters need to be determined at a physical resolution common to
all investigated clouds. Since most of our clouds are reasonably nearby, we
chose 0.1\,pc, which corresponds to the Jeans mass of a core with a temperature
of 15\,K and a density of GMCs of 5\,$\cdot$\,10$^4$\,cm$^{-3}$. 

There are in principle the following ways to determine the structure parameters
at this common scale for all clouds: i) determine the extinction maps for each
cloud at the appropriate spatial resolution; ii) determine the the maps for all
clouds at a sufficiently high spatial resolution and rebin them to the correct
resolution before performing the data analysis; iii) Determine the cloud
properties at a number of spatial resolutions and interpolate them to 0.1\,pc.
Option i) would certainly be the most desirable, however the most laborious. 
Below we will show that option ii) should be performed with great care, since it
can lead to erroneous results. Option iii) does not only allow us to determine
the cloud properties at the common physical scale, but also to investigate the
properties at other spatial resolutions.

In the left panel of Fig.\,\ref{Aur1FS} we show how the slope $\gamma_{\rm
low}$ for the cloud Auriga\,1 depends on the spatial resolution. At each
spatial resolution we determine the slope four times using different histogram
widths, as discussed in Sect.\,\ref{colden}. As one can see, these slopes agree
very well and can be averaged (solid line in the right panel of
Fig.\,\ref{Aur1FS}). For the other clouds a similar behavior is found (see
Appendix\,\ref{app1} for the remaining plots). About half the clouds show no
dependence of the slope on the scale and the other half a significant decrease
of $\gamma_{\rm low}$ with decreasing resolution.

The general trend for Auriga\,1 is that the slope $\gamma_{\rm low}$ decreases
towards larger spatial scales. On can understand this trend simply by assuming
that the cloud contains a number of very small, high extinction cores which are
simply not picked up at the larger spatial scales. This is not the case for all
clouds investigated. In some cases we find more or less constant slopes with
changing spatial scale. In those clouds the fraction of very small high
extinction cores is probably smaller.

We investigated what happens when we simply rebin the highest spatial resolution
image to the lower resolution and we re-determine the slopes $\gamma_{\rm low}$.
The results for Auriga\,1 are shown in the right panel in Fig.\,\ref{Aur1FS} as
a dotted line. One finds that the general trend of the slope values is retained.
However, individual $\gamma_{\rm low}$ values in the rebinned images can differ
by almost the one sigma uncertainties of the values obtained in the maps
determined at the respective spatial resolution. See Appendix\,\ref{app2} and
\ref{app3} for the effects the rebinning has on the $\gamma_{\rm low}$ and
$\gamma_{\rm high}$ values. In general the differences are well below the one
sigma uncertainties. However, in a few cases, such as the above quoted example
of Auriga\,1, larger differences can be found.

As a consequence of this result we chose to perform option iii) to determine our
cloud structure parameters for all clouds at 0.1\,pc resolution. We calculate
all parameters at each of the four spatial resolutions available to us, and then
interpolate to obtain the values at 0.1\,pc. All subsequent analysis is
performed this way.

\begin{table*}
\caption{\label{bintab} Fit parameters for the Auriga\,1 cloud obtained
for the various spatial resolutions and histogram bin sizes. We list the
parameters from the fit of the log-normal distribution, as well as the slopes
$\gamma$ and $\delta$.}
\centering
\begin{tabular}{cccrrrrrrrrr}
Sp. Res. & Sp. Res. & Bin size & $a$ & $A_0$ & $A_1$ & $\sigma$ & $rms$ & $\gamma_{\bf low}$ & $\gamma_{\bf high}$ & $\delta_{\bf low}$ & $\delta_{\bf high}$ \\
${\rm [}$arcmin${\rm ]}$ & [pc] & [mag] & [mag] & [mag] & [mag] & [mag] & [$\sigma$] & & & & \\ \hline 
0.5 & 0.065 & 0.125 & 26.2    & -24.2    & 25.0    & 1.06 &  6.4 & -0.49 & -0.22 & -0.38 & -0.14 \\
0.5 & 0.065 & 0.250 & 26.9    & -24.7    & 25.6    & 1.06 &  4.7 & -0.49 & -0.23 & -0.38 & -0.14 \\
0.5 & 0.065 & 0.500 & 31.5    & -29.2    & 30.1    & 1.05 &  3.4 & -0.49 & -0.23 & -0.38 & -0.14 \\
0.5 & 0.065 & 1.000 & 84.6    & -80.2    & 81.0    & 1.02 &  2.0 & -0.49 & -0.22 & -0.39 & -0.14 \\
1.0 & 0.131 & 0.125 & 11.1    & \,\,-9.8 & 10.8    & 1.12 &  8.8 & -0.50 & -0.26 & -0.36 & -0.13 \\
1.0 & 0.131 & 0.250 & 11.2    & \,\,-9.8 & 10.9    & 1.12 &  6.5 & -0.49 & -0.27 & -0.36 & -0.13 \\
1.0 & 0.131 & 0.500 & 11.5    & -10.1    & 11.5    & 1.12 &  4.8 & -0.49 & -0.26 & -0.37 & -0.13 \\
1.0 & 0.131 & 1.000 & 13.2    & -11.7    & 12.8    & 1.11 &  2.9 & -0.49 & -0.25 & -0.37 & -0.13 \\
2.0 & 0.262 & 0.125 & \,\,4.1 & \,\,-3.2 & \,\,4.1 & 1.23 & 12.2 & -0.56 & -0.33 & -0.36 & -0.14 \\
2.0 & 0.262 & 0.250 & \,\,4.1 & \,\,-3.2 & \,\,4.2 & 1.24 &  9.0 & -0.56 & -0.32 & -0.37 & -0.14 \\
2.0 & 0.262 & 0.500 & \,\,4.2 & \,\,-3.4 & \,\,4.3 & 1.23 &  6.6 & -0.56 & -0.32 & -0.38 & -0.14 \\
2.0 & 0.262 & 1.000 & \,\,5.5 & \,\,-4.1 & \,\,5.0 & 1.20 &  4.2 & -0.55 & -0.32 & -0.39 & -0.14 \\
4.0 & 0.524 & 0.125 & \,\,2.2 & \,\,-1.5 & \,\,2.3 & 1.32 & 16.0 & -0.69 & -0.45 & -0.47 & -0.27 \\
4.0 & 0.524 & 0.250 & \,\,2.2 & \,\,-1.6 & \,\,2.3 & 1.32 & 11.6 & -0.69 & -0.38 & -0.48 & -0.27 \\
4.0 & 0.524 & 0.500 & \,\,2.5 & \,\,-1.8 & \,\,2.5 & 1.29 &  9.3 & -0.70 & -0.41 & -0.48 & -0.28 \\
4.0 & 0.524 & 1.000 & \,\,3.4 & \,\,-2.5 & \,\,3.3 & 1.24 &  4.9 & -0.70 & -0.39 & -0.47 & -0.29 \\
\end{tabular}
\end{table*}

\subsection{Log-Normal Fits to the Column Density Distribution}
\label{lognorm}

Using the technique described in Sect.\,\ref{fittwo} we obtained fit parameters
for each cloud we analysed with Eq.\,\ref{lom12}. In Table\,\ref{bintab} we
summarise the fit parameters and the root mean square deviation ($rms$) values
obtained for the various spatial resolutions and histogram bin sizes for the
Auriga\,1 cloud as an example. There is a general trend visible for all
parameters. In particular the width of the distribution increases with spatial
scale. This is expected, since more and more small scale high extinction cores
are not detected anymore at these coarse resolutions. 

\begin{figure}
\centering
\includegraphics[width=8cm]{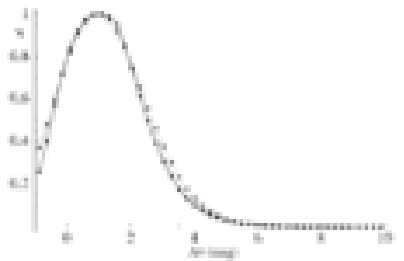} \\\
\includegraphics[width=8cm]{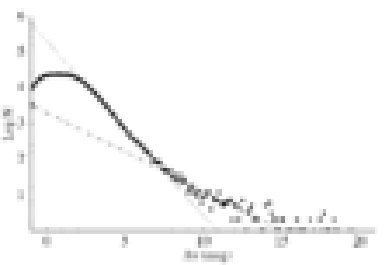} \\
\caption{\label{Aur1PBSF1} {\bf Top:} Plot of the best Log-Normal fit (dotted
line) to the normalised column density distribution (solid line) for the
Auriga\,1 cloud. The fit parameters are interpolated to a spatial scale of
0.1\,pc, and the data for the spatial resolution closest to 0.1\,pc are shown.
{\bf Bottom:} Plot of the $\log N$ vs $A_V$ column density distribution for the
Auriga\,1 cloud (symbols). Overplotted are the two fits obtained for the low and
high column density region.}
\end{figure}

In Fig.\,\ref{Aur1PBSF1} we show the normalised column density distribution for
the Auriga\,1 cloud as a solid line (shown is the data for the spatial
resolution closest to 0.1\,pc). Overplotted is a fit with parameters scaled to
0.1\,pc spatial scale. Similar plots for all individual clouds can be seen in
the Appendix\,\ref{app4}. We list the fit parameters and $rms$ values
(scaled to 0.1\,pc resolution) for all clouds in Table\,\ref{anatab}. 

 %  
 %  There is a correlation of the width ($\sigma$) of the distribution with the
 %  quality of the fit ($rms$). Thus, clouds that show a wider distribution of
 %  column densities can be less well fitted by a log-normal distribution. These
 %  are, in general, clouds that possess a significant excess of high $A_V$ regions
 %  compared to what would be expected from a turbulence driven log-normal
 %  distribution.
 %  

\begin{table*}
\caption{\label{anatab} Table of parameters of best fit for all clouds, at a
scale of 0.1\,pc. We list the parameters from the fit of the log-normal
distribution, the slopes $\gamma$ and $\delta$, as well as the cloud masses,
star formation threshold, extinction of a Jeans Mass core and the mass
fraction of clouds involved in star formation (MSF)}.
\centering
\begin{tabular}{lrrrrrrrrrrrrr}
Name & $a$ & $A_0$ & $A_1$ & $\sigma$ & $rms$ & $\gamma_{\rm low}$ & $\gamma_{\rm high}$ & $\delta_{\rm low}$ & 
$\delta_{\rm high}$ & $M_{\rm 1mag}$ & $A_{V, SF}$ & $A_{V, M_\odot}$ & $MSF$ \\
 & [mag] & [mag] & [mag] & [mag] & [$\sigma$] & & & & & [10$^3$M$_\odot$] & [mag] & [mag] & [\%] \\ \hline 
Auriga\,1        & 15.9    & -14.2    & 15.3    & 1.10 & 4.7 & -0.48 & -0.24 & -0.36 & -0.13 & 268 & 7.4 & 32 & 0.19\\ 
Auriga\,2        & \,\,5.5 & \,\,-4.6 & \,\,5.6 & 1.14 & 2.5 & -0.71 & -0.32 & -0.51 & -0.21 & 13  & 4.9 & 15 & 0.33\\ 
Cepheus          & 32.9    & -31.8    & 32.8    & 1.04 & 9.0 & -0.50 & -0.23 & -0.37 & -0.15 & 256 & 6.7 & 29 & 0.26\\ 
Chamaeleon       & \,\,3.3 & \,\,-2.7 & \,\,3.4 & 1.27 & 3.2 & -0.39 & -0.18 & -0.25 & -0.13 & 45  & 7.3 & 24 & 0.81\\ 
Circinus         & \,\,9.8 & \,\,-7.5 & \,\,9.9 & 1.17 & 4.5 & -0.36 & -0.17 & -0.24 & -0.16 & 113 & 6.8 & 30 & 1.4 \\ 
Corona Australis & \,\,2.2 & \,\,-1.7 & \,\,2.5 & 1.46 & 1.6 & -0.40 & -0.16 & -0.20 & -0.15 & 1   & 3.7 & 23 & 9.6 \\ 
$\lambda$-Ori    & 12.8    & -12.3    & 12.9    & 1.11 & 4.6 & -0.51 & -0.34 & -0.40 & -0.13 & 122 & 7.2 & 28 & 0.12\\ 
Lupus\,1, 2      & \,\,8.0 & \,\,-7.4 & \,\,8.0 & 1.08 & 3.4 & -0.73 & -0.20 & -0.47 & -0.13 & 4.6 & 3.4 & 23 & 2.4 \\ 
Lupus\,3, 4, 5, 6& \,\,4.2 & \,\,-2.7 & \,\,4.3 & 1.22 & 2.0 & -0.73 & -0.21 & -0.47 & -0.12 & 24  & 5.1 & 25 & 0.40\\ 
Monoceros        & 12.5    & -11.9    & 12.6    & 1.11 & 6.7 & -0.42 & -0.18 & -0.32 & -0.12 & 74  & 4.8 & 35 & 2.0 \\ 
Ophiuchus        & \,\,1.5 & \,\,-0.3 & \,\,1.5 & 1.48 & 7.0 & -0.27 & -0.11 & -0.21 & -0.13 & 9.3 & 8.6 & 27 & 0.78\\ 
Orion\,A         & \,\,3.9 & \,\,-3.0 & \,\,4.1 & 1.39 & 4.2 & -0.28 & -0.15 & -0.19 & -0.12 & 44  & 5.4 & 37 & 5.0 \\ 
Orion\,B         & \,\,6.4 & \,\,-5.5 & \,\,6.5 & 1.23 & 4.3 & -0.39 & -0.13 & -0.26 & -0.11 & 78  & 6.8 & 38 & 1.1 \\ 
Perseus          & \,\,5.8 & \,\,-5.0 & \,\,5.9 & 1.23 & 3.8 & -0.42 & -0.22 & -0.28 & -0.17 & 29  & 4.8 & 25 & 3.0 \\ 
Serpens          & 12.3    & \,\,-7.3 & 12.4    & 1.13 & 2.1 & -0.32 & -0.16 & -0.21 & -0.14 & 18  & 7.7 & 31 & 1.3 \\ 
Taurus           & \,\,4.5 & \,\,-3.8 & \,\,4.6 & 1.26 & 3.2 & -0.34 & -0.15 & -0.24 & -0.14 & 19  & 4.4 & 28 & 4.8 \\
\end{tabular}                                                                                             
\end{table*}                                                                                              
 
\subsection{The $\log(N)$ vs $A_V$ Column Density Distribution}
\label{strght}

As described in Sect.\,\ref{fitone} we calculate gradients $\gamma$ for each
cloud in our sample. As discussed, there are usually at least two distinct
regions with different slopes. One region, at low extinction values
($\gamma_{\rm low}$) characterises the general turbulence of the cloud. At
higher extinction values ($\gamma_{\rm high}$) gravity becomes important and
changes the column density away from a log-normal distribution. As an example we
show the $\log(N)$ vs $A_V$  for Auriga\,1 in Fig.\,\ref{Aur1PBSF1}. The plots
for the other clouds are shown in Appendix\,\ref{app5}. We also show how the
slopes $\gamma_{\rm low}$ and $\gamma_{\rm high}$ depend on the spatial
resolution for all clouds in Fig.\,\ref{allSG}. 

The gradients for all clouds, averaged and interpolated to 0.1\,pc
resolution, are summarised in Table\,\ref{anatab}. In general the $\gamma_{\rm
low}$ values are about twice as negative as the $\gamma_{\rm high}$ values. More
importantly the scatter for the slopes in the high column density regions of the
clouds is smaller than in the low $A_V$ regions. For the averages and scatter
for all clouds we find: $\left< \gamma_{\rm low} \right> = -0.45 \pm 0.15$ and
$\left< \gamma_{\rm high} \right> = -0.20 \pm 0.06$. This indicates, that once
gravity becomes important enough to influence the column density distribution
(i.e. star formation starts), then we will find far fewer differences between
the various clouds. While at low column densities, where external factors
(proximity to Supernovae, etc.) determine the turbulent motions, much larger
cloud to cloud differences are seen.

\begin{figure}
\centering
\includegraphics[width=8cm]{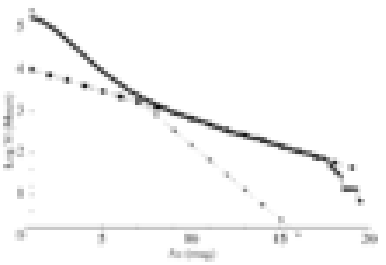} \\
\caption{\label{aur1mass} Plot of the mass distribution in the cloud Auriga\,1
together with the two fits to the low and high column density regime.} 
\end{figure}

\subsection{The $\log(M)$ vs $A_V$ Mass Distribution}
\label{massres}

As for the column density distribution we determine the slopes $\delta$ of the
$\log(M)$ vs $A_V$ mass distribution for low and high column densities. We show
as an example the mass distribution of the Auriga\,1 cloud in
Fig.\,\ref{aur1mass}. Overplotted are the linear fits. The values for all
slopes $\delta$ for different spatial resolutions are listed in
Table\,\ref{bintab}. 

\begin{figure*}
\centering
\includegraphics[width=8cm]{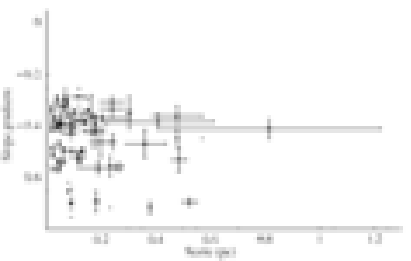} \hfill
\includegraphics[width=8cm]{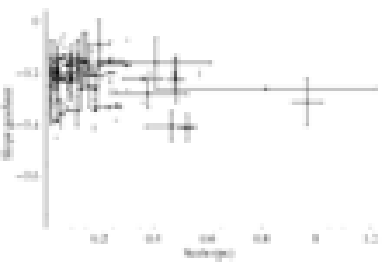} \\
\includegraphics[width=8cm]{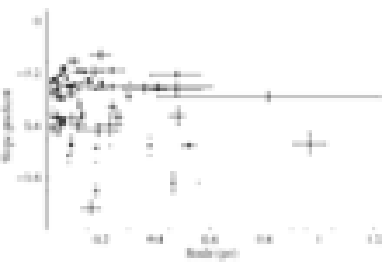} \hfill
\includegraphics[width=8cm]{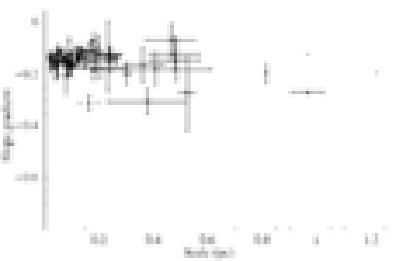} \\
\caption{\label{allSG} {\bf Top Left:} A plot showing measured slope gradients
$\gamma_{\rm low}$ against scale for all the clouds studied. {\bf Top Right:} As
in the top left panel but for the high $A_V$ region. {\bf Bottom panels:} As in
the top panels but for the slope gradients $\delta$ of the mass distribution in
the clouds. All plots are shown to the same scale to show the differences in the
scatter.} 
\end{figure*}

The slopes for all clouds averaged and interpolated to 0.1\,pc are
summarised in Table\,\ref{anatab}, the mass distributions are shown in
Appendix\,\ref{app6}. As for the values of $\gamma$, the scatter of the slopes
differs for the low and high column density material. For the averages of all
clouds we find: $\left< \delta_{\rm low} \right> = -0.21 \pm 0.11$ and $\left<
\delta_{\rm high} \right> = -0.14 \pm 0.025$. Again, we find that the star
forming (high column density) parts of the clouds are very similar, while the
low $A_V$ (turbulence dominated) regions show a larger scatter. We show how the
slopes $\delta_{\rm low}$ and $\delta_{\rm high}$ depend on the spatial
resolution for all clouds in Fig.\,\ref{allSG}. 

We extrapolate the fit to the low $A_V$ regions, which leads to the total mass
of the cloud. For the purpose of this paper, we calculate the mass in the cloud
at a column density of above 1\,mag of optical extinction ($M_{\rm 1mag}$). This
ensures we do not include noise and also only integrate over material that is
above the threshold for self-shielding from UV radiation and molecular hydrogen
formation (Hartmann et al. \cite{2001ApJ...562..852H}). Extrapolating the high
$A_V$ fit towards one solar mass, one can find the extinction value a core with
the Jeans Mass would have ($A_{V, M_\odot}$). Hence, this characterises the
maximum extinction in the cloud before collapse starts. Furthermore, we
determine the intercept between the low and high $A_V$ region, to characterise
above which column density ($A_{V, SF}$) material is more likely involved in
star formation than being characterised by the turbulence in the cloud. This can
be seen as the star formation threshold, i.e. the column density above which
star formation occurs in the cloud. Finally, the ratio of the mass in the cloud
above an extinction $A_{V, SF}$ (the mass currently associated with star
formation) to the total mass of the cloud above 1\,mag $A_V$ is determined.
Assuming that about 1/3 of the total mass associated with star formation is
transformed into stars (Alves et al. \cite{2007A&A...462L..17A}) we can estimate
the overall fraction of mass involved in star formation ($MSF$) an
indicator of the star formation efficiency. All these values for each cloud are
summarised in Table\,\ref{anatab}.

\section{Discussion and Conclusions}
\label{dissc}

We investigated the column density and mass distribution of a number of GMCs
based on near infrared median colour excess extinction maps determined from
2MASS. Our multi scale extinction mapping approach enabled us to determine the
cloud properties homogeneously at a common scale of 0.1\,pc for all clouds.
Furthermore, it allowed us to investigate how the cloud properties depend on the
spatial scale.

We find that for about half the investigated clouds the slope of the column
density and mass distribution changes to more negative (steeper) values with
increasing spatial scale. For the other half no change is found over a scale
range of almost an order of magnitude. The former can be understood by the fact
that at larger scales, small scale high extinction cores are not detected
anymore, naturally leading to a steeper mass and column density distribution. In
the latter case this does not happen. This could mean that there are no small
scale high extinction regions. But this does not seem plausible, in particular
since the clouds in this group are e.g. Taurus, Perseus, Ophiuchus and Orion.
Another possibility is that there is no significant structure in those clouds at
the smallest scales, and hence no change over the observed range of scales. We
will analyse this in more detail in a forthcoming paper where we will determine
the structure functions of all clouds. 

Our fits to the column density distributions using a log-normal function
resulted in variable outcomes. Some clouds, in particular Auriga\,2, Corona
Australis, Lupus\,3, 4, 5, 6 and Serpens can be fit very well ($rms$ lower than
3\,$\sigma$). Other clouds (Cepheus, Monoceros, Ophiuchus) cannot be fit
properly by a log-normal function. This is in agreement with earlier works on
some of these clouds e.g. by Lombardi et al. \cite{2008A&A...489..143L} and more
recently by Kainulainen et al. \cite{2009A&A...........K}. All investigated
clouds show an excess of column density compared to a log-normal distribution at
higher $A_V$ values. The quality of the fit is hence a measure of how much
excess there is in a particular cloud. This excess material has decoupled from
the general turbulent field and is dominated by gravity, in other words more
likely actively involved in star formation. However, in our sample we could not
find any significant correlations of the achieved quality of the log-normal fit
with other cloud properties or the amount of young stars in the cloud.

The analysis of the mass distribution enables us to draw further conclusions
about the structure of the clouds and their star formation properties. For all
clouds the change in the slope of the mass distribution is more pronounced than
in the column density distribution. It is hence easier to determine the
threshold at which gravity becomes the dominant force in shaping the structure.
This value ranges from 3.4\,mag in Lupus\,1, 2 to 8.6\,mag of optical extinction
in Ophiuchus. We associate this with the extinction threshold for star formation,
originally found to be about 7\,mag of $A_V$ in Ophiuchus by Johnstone et al.
\cite{2004ApJ...611L..45J}, in agreement with our value. On average for all
clouds we find $\left< A_{\rm V, SF} \right> = 6.0 \pm 1.5$\,mag optical
extinction, a rather small range.

In contrast, the mass fraction of each cloud which is currently involved in star
formation varies by almost two orders of magnitude. It ranges from 0.12\,\% in
$\lambda$-Ori to almost 10\,\% in Corona Australis. We note that these values
are not an estimate of the star formation efficiency in these clouds. They
rather should be seen as the potential of the cloud to form new stars in the
next few 10$^6$\,yrs. The wide range of values indicates that our sample
contains clouds at different stages in their evolution. We see clouds that
currently only form a small number of stars (e.g. Auriga\,1 and Corona
Australis) but with a completely different fraction of mass involved in star
formation. In Auriga\,1 only a small fraction of the total material seems to be
available for future star formation. On the contrary in Corona Australis a much
larger fraction of material is expected to be forming stars. While other regions
that currently form stars (Orion, Taurus, Perseus) still possess a significant
fraction of material in a state which is expected to continue star formation.
One example for a cloud which might have reached the end of star formation seems
to be Ophiuchus. It is currently forming a large number of stars, but has only
less than one percent of material in a gravity dominated form. 

We have searched amongst our determined cloud properties for further
correlations. In particular we hoped to find a link between the predominant mode
of star formation in the cloud (clustered vs. isolated) with one or several
cloud properties. No such correlation could be found. This is partly expected
since our extinction maps show only how much material is potentially involved in
star formation, and not how much gas and dust (and its properties) have lead to
the star formation mode we currently see in these clouds. One could expect that
the properties of the turbulence dominated part of the clouds (out of which the
denser parts are formed) show some dependence on the star formation mode. Either
because different turbulent properties cause different modes of star formation,
or that feedback from young star clusters and isolated YSOs causes different
turbulent properties. But again, no such dependence could be found in our data.

However, there are clear differences in the properties of the clouds at low
column densities compared to high $A_V$ values. The scatter of the $\gamma_{\rm
low}$ and $\delta_{\rm low}$ values between clouds is a factor of 2.5 (for
$\gamma$) and 4.5 (for $\delta$) larger than for the high extinction regions.
Essentially this paints the following picture of the cloud structure: The low
$A_V$ and turbulence dominated regions differ from cloud to cloud. Their column
density and mass distributions are determined by the environment of the cloud
and the feedback they have experienced from the star formation processes within
them. Above a (column) density threshold of about 6\,mag $A_V$, which is
independent of the cloud, gravity becomes the dominant force in shaping the
structure. This part of the cloud is then more and more decoupled from the
influences of its surrounding turbulent field, and thus the column density and
mass distributions for all clouds are virtually identical above the extinction
threshold. This also implies that local feedback from young clusters or stars
has no significant influence in shaping the column density and mass
distributions.

\section*{acknowledgements}
\label{acks}

JR acknowledges a University of Kent scholarship. This publication makes use of
data products from the Two Micron All Sky  Survey, which is a joint project of
the University of Massachusetts and the Infrared Processing and Analysis
Center/California Institute  of Technology, funded by the National Aeronautics
and Space  Administration and the National Science Foundation.

\label{lastpage}

 %  
 %  \clearpage
 %  \newpage
 %  
 %  \begin{appendix}
 %  
 %  \section{Dependence of $\gamma_{\rm low}$ on scale}
 %  \label{app1}
 %  \label{app2}
 %  \label{app3}
 %  \label{app4}
 %  \label{app5}
 %  \label{app6}
 %  
 %  \clearpage
 %  \newpage
 %  
 %  \begin{figure}
 %  \centering
 %  \includegraphics[width=8.5cm]{./images/FS_Auriga1.eps}
 %  \caption{\label{app1_1} Dependence of the slope $\gamma_{\rm low}$ for the
 %  cloud Auriga\,1 on the spatial scale. Shown are the values for each spatial
 %  resolution and histogram bin width.}
 %  \end{figure}
 %  
 %  \end{appendix}
 %  \end{document}
 %  

\begin{appendix}

\section{Dependence of $\gamma_{\rm low}$ on scale}
\label{app1}

\clearpage
\newpage

\begin{figure}
\centering
\includegraphics[width=8.5cm]{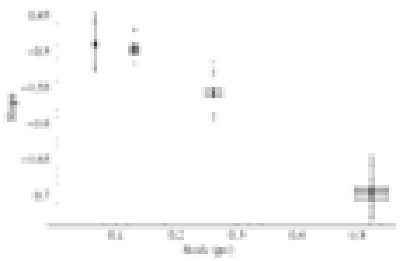}
\caption{\label{app1_1} Dependence of the slope $\gamma_{\rm low}$ for the
cloud Auriga\,1 on the spatial scale. Shown are the values for each spatial
resolution and histogram bin width.}
\end{figure}

\begin{figure}
\centering
\includegraphics[width=8.5cm]{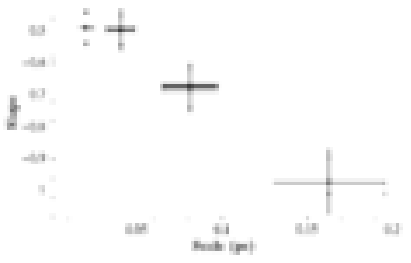}
\caption{As Fig.\,\ref{app1_1} but for the cloud Auriga\,2.}
\end{figure}

\begin{figure}
\centering
\includegraphics[width=8.5cm]{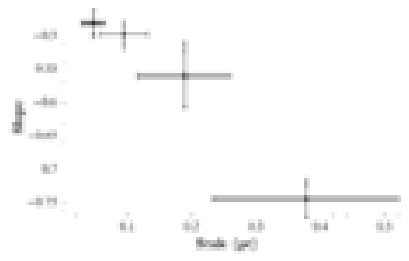}
\caption{As Fig.\,\ref{app1_1} but for the cloud Cepheus.}
\end{figure}

\begin{figure}
\centering
\includegraphics[width=8.5cm]{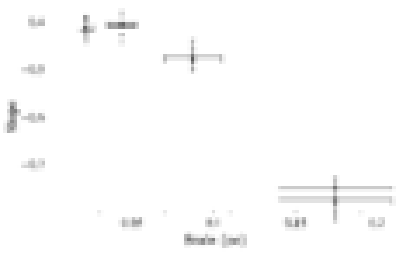}
\caption{As Fig.\,\ref{app1_1} but for the cloud Chamaeleon.}
\end{figure}

\begin{figure}
\centering
\includegraphics[width=8.5cm]{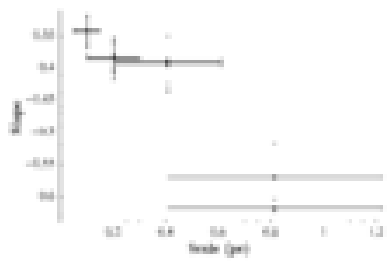}
\caption{As Fig.\,\ref{app1_1} but for the cloud Circinus.}
\end{figure}

\begin{figure}
\centering
\includegraphics[width=8.5cm]{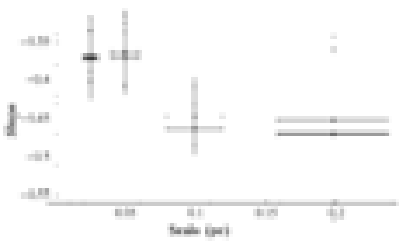}
\caption{As Fig.\,\ref{app1_1} but for the cloud Corona Australis.}
\end{figure}

\begin{figure}
\centering
\includegraphics[width=8.5cm]{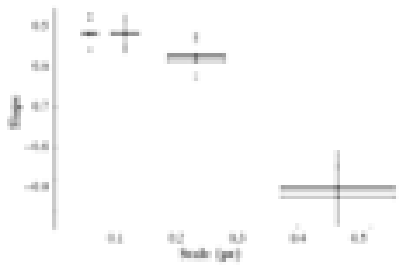}
\caption{As Fig.\,\ref{app1_1} but for the cloud $\lambda$-Ori.}
\end{figure}

\begin{figure}
\centering
\includegraphics[width=8.5cm]{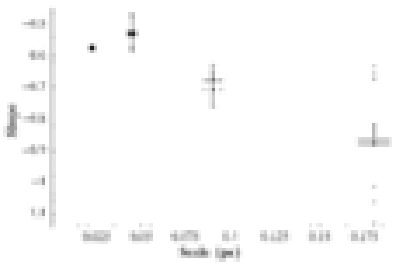}
\caption{As Fig.\,\ref{app1_1} but for the cloud Lupus\,1, 2.}
\end{figure}

\begin{figure}
\centering
\includegraphics[width=8.5cm]{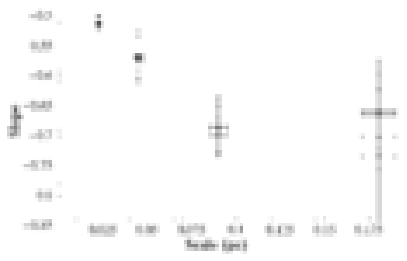}
\caption{As Fig.\,\ref{app1_1} but for the cloud Lupus\,3, 4, 5, 6.}
\end{figure}

\begin{figure}
\centering
\includegraphics[width=8.5cm]{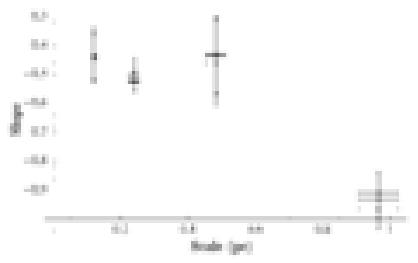}
\caption{As Fig.\,\ref{app1_1} but for the cloud Monoceros.}
\end{figure}

\begin{figure}
\centering
\includegraphics[width=8.5cm]{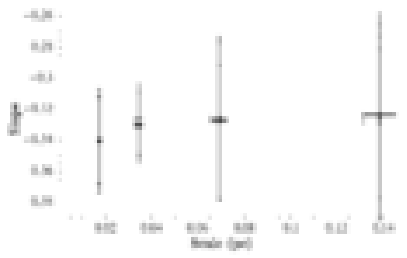}
\caption{As Fig.\,\ref{app1_1} but for the cloud Ophiuchus.}
\end{figure}

\begin{figure}
\centering
\includegraphics[width=8.5cm]{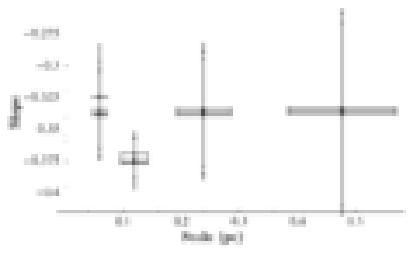}
\caption{As Fig.\,\ref{app1_1} but for the cloud Orion\,A.}
\end{figure}

\begin{figure}
\centering
\includegraphics[width=8.5cm]{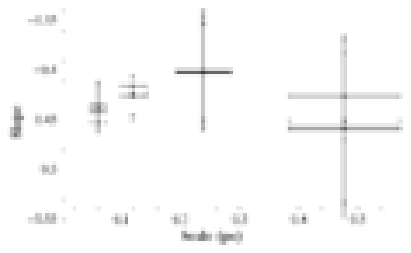}
\caption{As Fig.\,\ref{app1_1} but for the cloud Orion\,B.}
\end{figure}

\begin{figure}
\centering
\includegraphics[width=8.5cm]{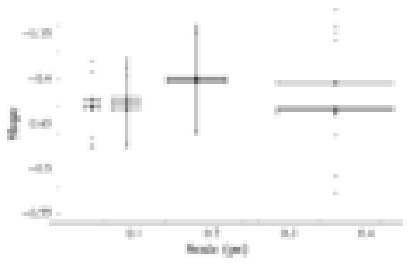}
\caption{As Fig.\,\ref{app1_1} but for the cloud Perseus.}
\end{figure}

\begin{figure}
\centering
\includegraphics[width=8.5cm]{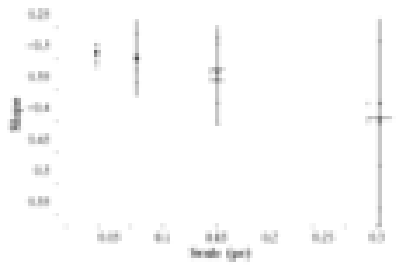}
\caption{As Fig.\,\ref{app1_1} but for the cloud Serpens.}
\end{figure}

\begin{figure}
\centering
\includegraphics[width=8.5cm]{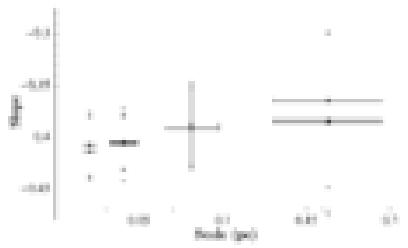}
\caption{As Fig.\,\ref{app1_1} but for the cloud Taurus.}
\end{figure}

\clearpage
\newpage

\subsection{Effect of rebinning of data on $\gamma_{\rm low}$}
\label{app2}

\clearpage
\newpage

\begin{figure}
\centering
\includegraphics[width=8.5cm]{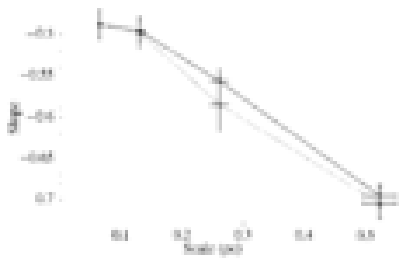}
\caption{\label{app2_1} Dependence of the slope $\gamma_{\rm low}$ for the
cloud Auriga\,1 on the spatial scale for the original data (solid line) and
when the highest resolution image is rebinned to the lower spatial resolution
(dotted line).}
\end{figure}

\begin{figure}
\centering
\includegraphics[width=8.5cm]{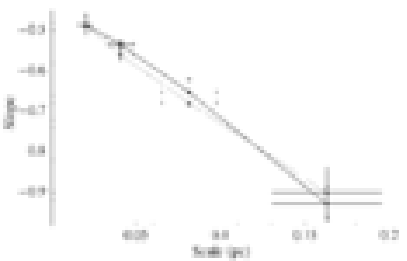}
\caption{As Fig.\,\ref{app2_1} but for the cloud Auriga\,2.}
\end{figure}

\begin{figure}
\centering
\includegraphics[width=8.5cm]{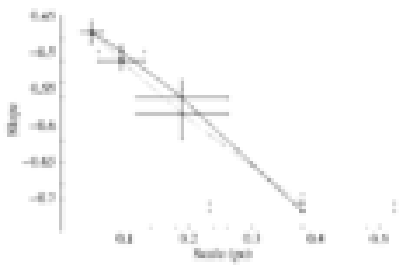}
\caption{As Fig.\,\ref{app2_1} but for the cloud Cepheus.}
\end{figure}

\begin{figure}
\centering
\includegraphics[width=8.5cm]{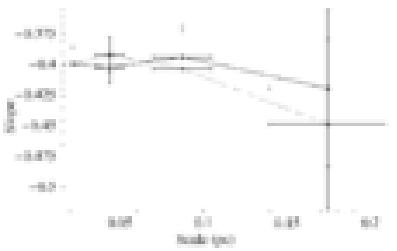}
\caption{As Fig.\,\ref{app2_1} but for the cloud Chamaeleon.}
\end{figure}

\begin{figure}
\centering
\includegraphics[width=8.5cm]{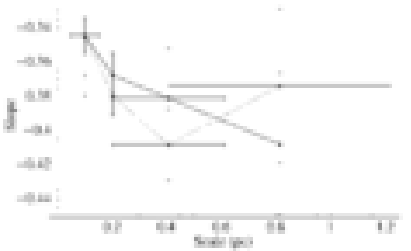}
\caption{As Fig.\,\ref{app2_1} but for the cloud Circinus.}
\end{figure}

\begin{figure}
\centering
\includegraphics[width=8.5cm]{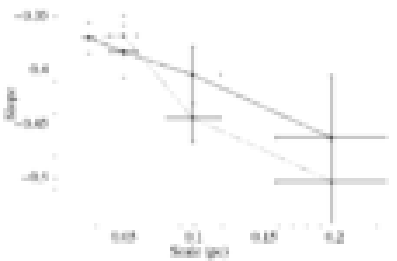}
\caption{As Fig.\,\ref{app2_1} but for the cloud Corona Australis.}
\end{figure}

\begin{figure}
\centering
\includegraphics[width=8.5cm]{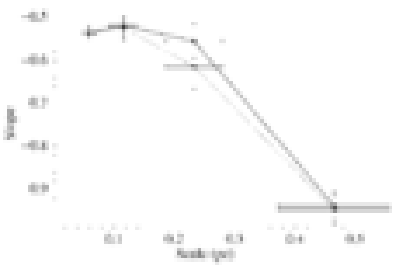}
\caption{As Fig.\,\ref{app2_1} but for the cloud $\lambda$-Ori.}
\end{figure}

\begin{figure}
\centering
\includegraphics[width=8.5cm]{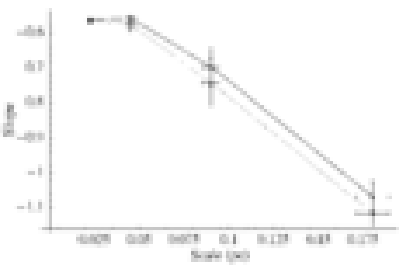}
\caption{As Fig.\,\ref{app2_1} but for the cloud Lupus\,1, 2.}
\end{figure}

\begin{figure}
\centering
\includegraphics[width=8.5cm]{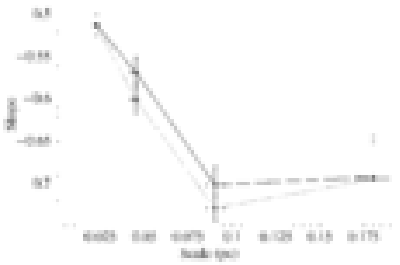}
\caption{As Fig.\,\ref{app2_1} but for the cloud Lupus\,3, 4, 5, 6.}
\end{figure}

\begin{figure}
\centering
\includegraphics[width=8.5cm]{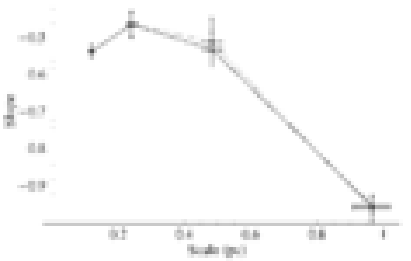}
\caption{As Fig.\,\ref{app2_1} but for the cloud Monoceros.}
\end{figure}

\begin{figure}
\centering
\includegraphics[width=8.5cm]{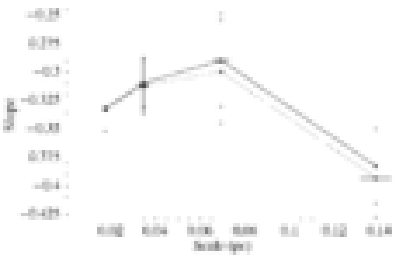}
\caption{As Fig.\,\ref{app2_1} but for the cloud Ophiuchus.}
\end{figure}

\begin{figure}
\centering
\includegraphics[width=8.5cm]{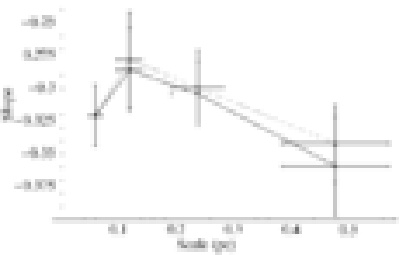}
\caption{As Fig.\,\ref{app2_1} but for the cloud Orion\,A.}
\end{figure}

\begin{figure}
\centering
\includegraphics[width=8.5cm]{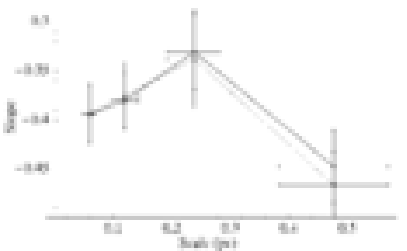}
\caption{As Fig.\,\ref{app2_1} but for the cloud Orion\,B.}
\end{figure}

\begin{figure}
\centering
\includegraphics[width=8.5cm]{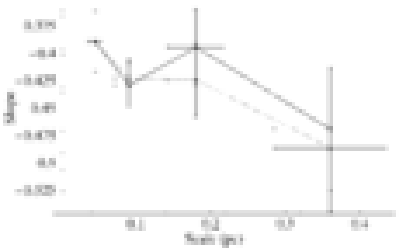}
\caption{As Fig.\,\ref{app2_1} but for the cloud Perseus.}
\end{figure}

\begin{figure}
\centering
\includegraphics[width=8.5cm]{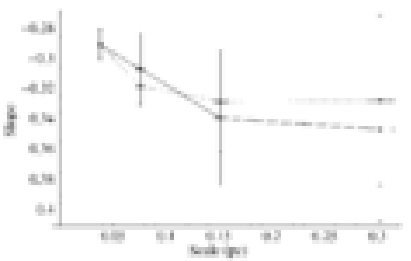}
\caption{As Fig.\,\ref{app2_1} but for the cloud Serpens.}
\end{figure}

\begin{figure}
\centering
\includegraphics[width=8.5cm]{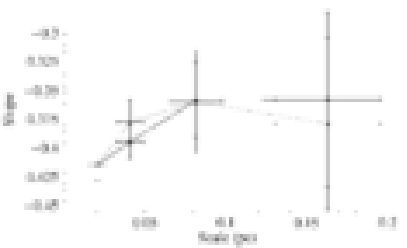}
\caption{As Fig.\,\ref{app2_1} but for the cloud Taurus.}
\end{figure}

\clearpage
\newpage

\subsection{Effect of rebinning of data on $\gamma_{\rm high}$}
\label{app3}

\clearpage
\newpage

\begin{figure}
\centering
\includegraphics[width=8.5cm]{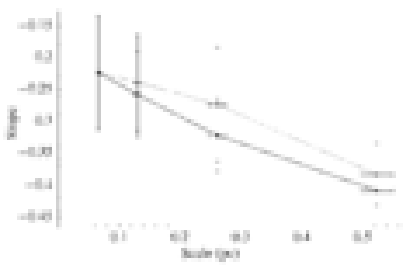}
\caption{\label{app3_1} Dependence of the slope $\gamma_{\rm high}$ for the
cloud Auriga\,1 on the spatial scale for the original data (solid line) and
when the highest resolution image is rebinned to the lower spatial resolution
(dotted line).}
\end{figure}

\begin{figure}
\centering
\includegraphics[width=8.5cm]{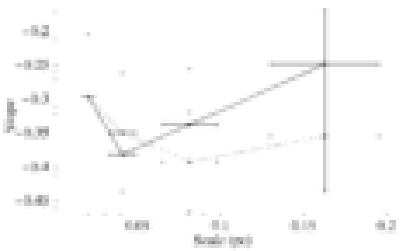}
\caption{As Fig.\,\ref{app3_1} but for the cloud Auriga\,2.}
\end{figure}

\begin{figure}
\centering
\includegraphics[width=8.5cm]{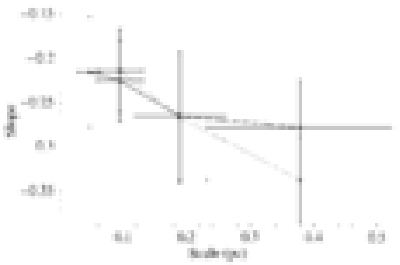}
\caption{As Fig.\,\ref{app3_1} but for the cloud Cepheus.}
\end{figure}

\begin{figure}
\centering
\includegraphics[width=8.5cm]{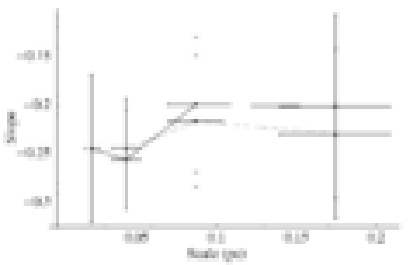}
\caption{As Fig.\,\ref{app3_1} but for the cloud Chamaeleon.}
\end{figure}

\begin{figure}
\centering
\includegraphics[width=8.5cm]{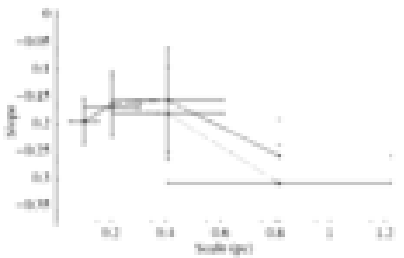}
\caption{As Fig.\,\ref{app3_1} but for the cloud Circinus.}
\end{figure}

\begin{figure}
\centering
\includegraphics[width=8.5cm]{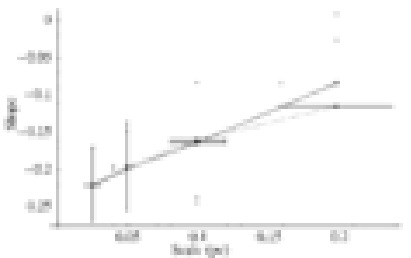}
\caption{As Fig.\,\ref{app3_1} but for the cloud Corona Australis.}
\end{figure}

\begin{figure}
\centering
\includegraphics[width=8.5cm]{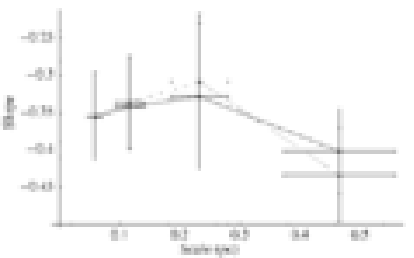}
\caption{As Fig.\,\ref{app3_1} but for the cloud $\lambda$-Ori.}
\end{figure}

\begin{figure}
\centering
\includegraphics[width=8.5cm]{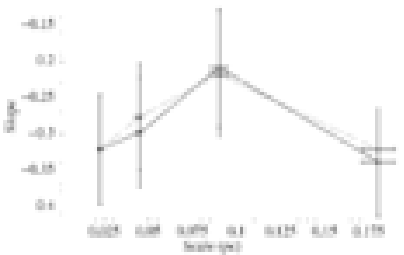}
\caption{As Fig.\,\ref{app3_1} but for the cloud Lupus\,1, 2.}
\end{figure}

\begin{figure}
\centering
\includegraphics[width=8.5cm]{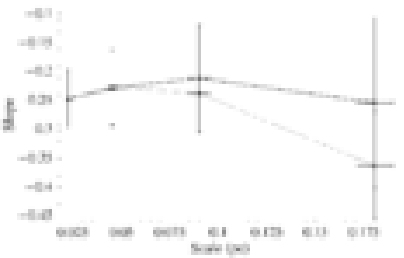}
\caption{As Fig.\,\ref{app3_1} but for the cloud Lupus\,3, 4, 5, 6.}
\end{figure}

\begin{figure}
\centering
\includegraphics[width=8.5cm]{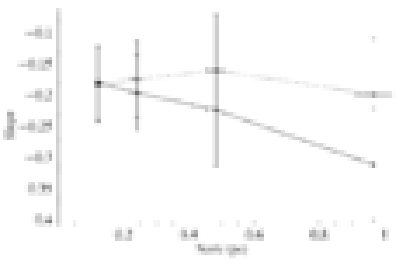}
\caption{As Fig.\,\ref{app3_1} but for the cloud Monoceros.}
\end{figure}

\begin{figure}
\centering
\includegraphics[width=8.5cm]{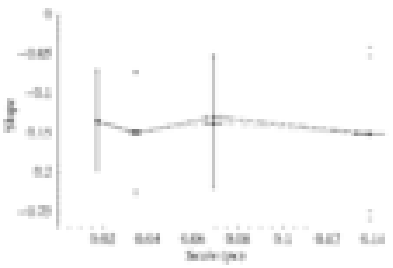}
\caption{As Fig.\,\ref{app3_1} but for the cloud Ophiuchus.}
\end{figure}

\begin{figure}
\centering
\includegraphics[width=8.5cm]{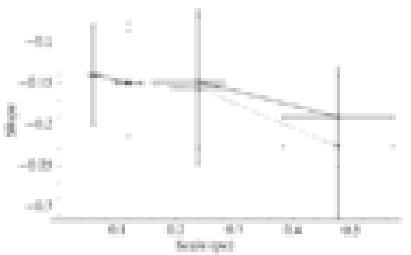}
\caption{As Fig.\,\ref{app3_1} but for the cloud Orion\,A.}
\end{figure}

\begin{figure}
\centering
\includegraphics[width=8.5cm]{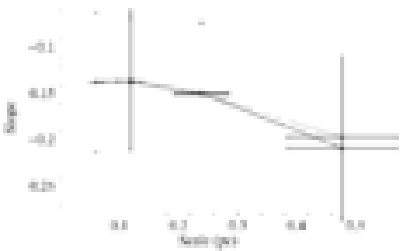}
\caption{As Fig.\,\ref{app3_1} but for the cloud Orion\,B.}
\end{figure}

\begin{figure}
\centering
\includegraphics[width=8.5cm]{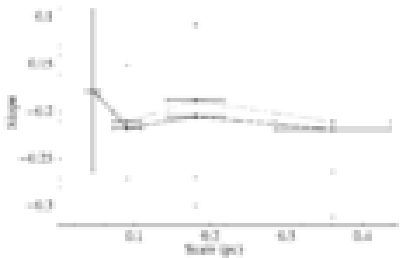}
\caption{As Fig.\,\ref{app3_1} but for the cloud Perseus.}
\end{figure}

\begin{figure}
\centering
\includegraphics[width=8.5cm]{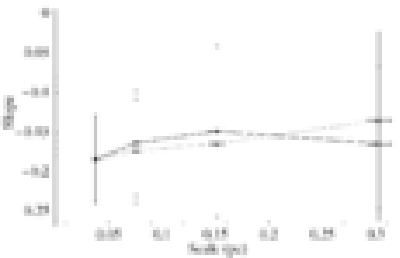}
\caption{As Fig.\,\ref{app3_1} but for the cloud Serpens.}
\end{figure}

\begin{figure}
\centering
\includegraphics[width=8.5cm]{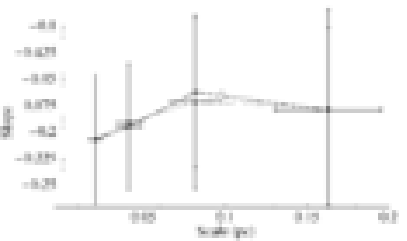}
\caption{As Fig.\,\ref{app3_1} but for the cloud Taurus.}
\end{figure}

\clearpage
\newpage

\subsection{Log-normal fits to the normalised Column Density Distributions}
\label{app4}

\clearpage
\newpage

\begin{figure}
\centering
\includegraphics[width=8.5cm]{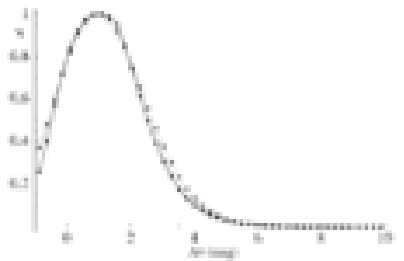}
\caption{\label{app4_1} Best log-normal fit (dotted line) to the Column Density
Distribution for the cloud Auriga\,1 (solid line) for a spatial scale of
0.1\,pc.}
\end{figure}

\begin{figure}
\centering
\includegraphics[width=8.5cm]{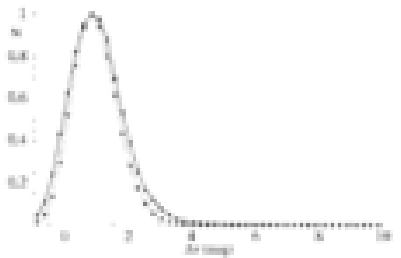}
\caption{As Fig.\,\ref{app4_1} but for the cloud Auriga\,2.}
\end{figure}

\begin{figure}
\centering
\includegraphics[width=8.5cm]{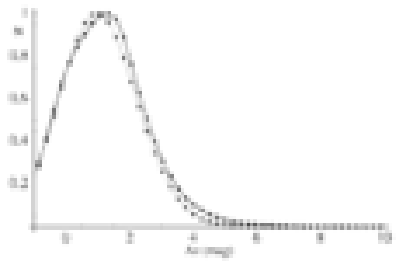}
\caption{As Fig.\,\ref{app4_1} but for the cloud Cepheus.}
\end{figure}

\begin{figure}
\centering
\includegraphics[width=8.5cm]{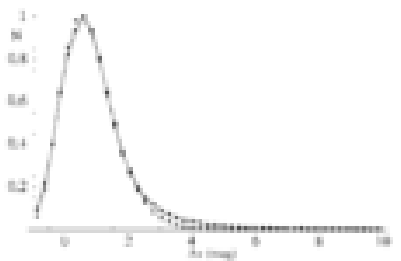}
\caption{As Fig.\,\ref{app4_1} but for the cloud Chamaeleon.}
\end{figure}

\begin{figure}
\centering
\includegraphics[width=8.5cm]{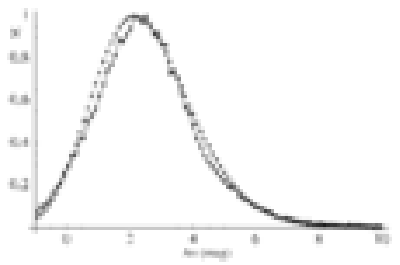}
\caption{As Fig.\,\ref{app4_1} but for the cloud Circinus.}
\end{figure}

\begin{figure}
\centering
\includegraphics[width=8.5cm]{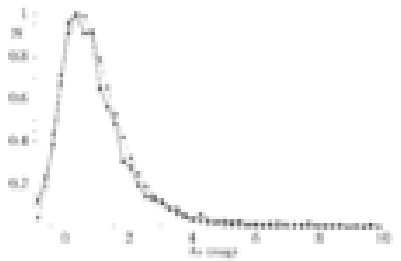}
\caption{As Fig.\,\ref{app4_1} but for the cloud Corona Australis.}
\end{figure}

\begin{figure}
\centering
\includegraphics[width=8.5cm]{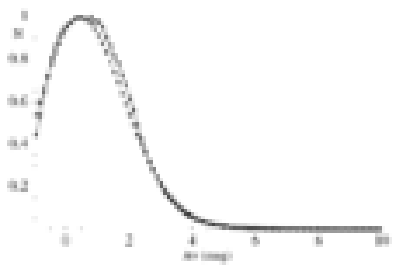}
\caption{As Fig.\,\ref{app4_1} but for the cloud $\lambda$-Ori.}
\end{figure}

\begin{figure}
\centering
\includegraphics[width=8.5cm]{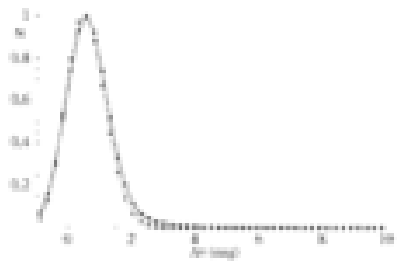}
\caption{As Fig.\,\ref{app4_1} but for the cloud Lupus\,1, 2.}
\end{figure}

\begin{figure}
\centering
\includegraphics[width=8.5cm]{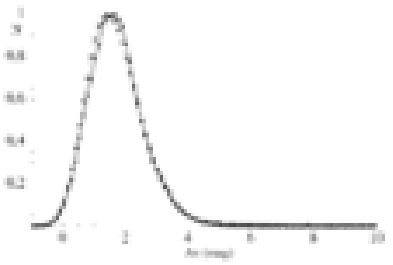}
\caption{As Fig.\,\ref{app4_1} but for the cloud Lupus\,3, 4, 5, 6.}
\end{figure}

\begin{figure}
\centering
\includegraphics[width=8.5cm]{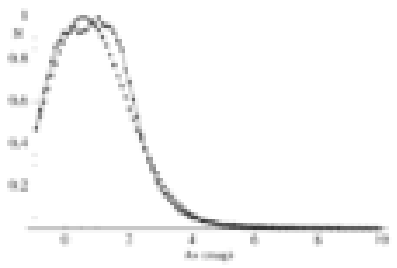}
\caption{As Fig.\,\ref{app4_1} but for the cloud Monoceros.}
\end{figure}

\begin{figure}
\centering
\includegraphics[width=8.5cm]{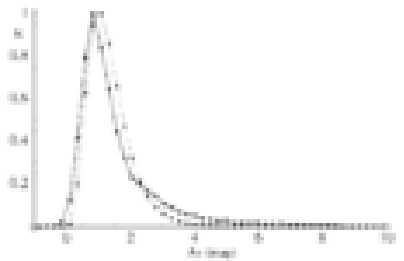}
\caption{As Fig.\,\ref{app4_1} but for the cloud Ophiuchus.}
\end{figure}

\begin{figure}
\centering
\includegraphics[width=8.5cm]{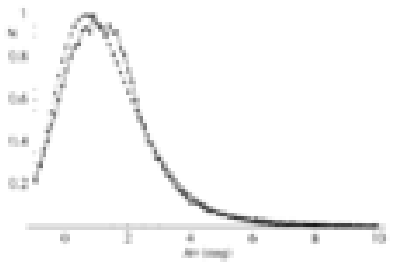}
\caption{As Fig.\,\ref{app4_1} but for the cloud Orion\,A.}
\end{figure}

\begin{figure}
\centering
\includegraphics[width=8.5cm]{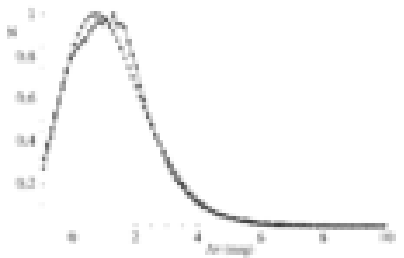}
\caption{As Fig.\,\ref{app4_1} but for the cloud Orion\,B.}
\end{figure}

\begin{figure}
\centering
\includegraphics[width=8.5cm]{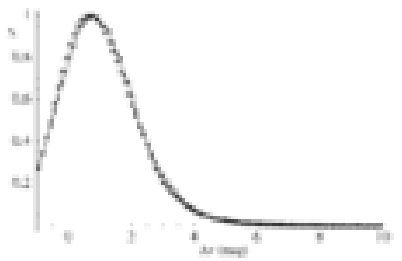}
\caption{As Fig.\,\ref{app4_1} but for the cloud Perseus.}
\end{figure}

\begin{figure}
\centering
\includegraphics[width=8.5cm]{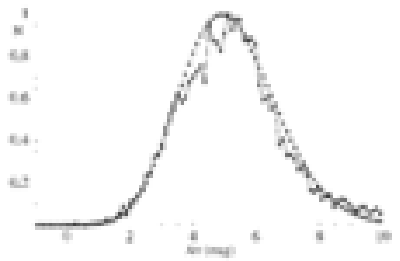}
\caption{As Fig.\,\ref{app4_1} but for the cloud Serpens.}
\end{figure}

\begin{figure}
\centering
\includegraphics[width=8.5cm]{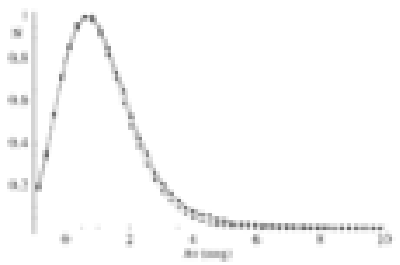}
\caption{As Fig.\,\ref{app4_1} but for the cloud Taurus.}
\end{figure}

\clearpage
\newpage

\subsection{The log(N) vs $A_V$ Column Density Distributions}
\label{app5}

\clearpage
\newpage

\begin{figure}
\centering
\includegraphics[width=8.5cm]{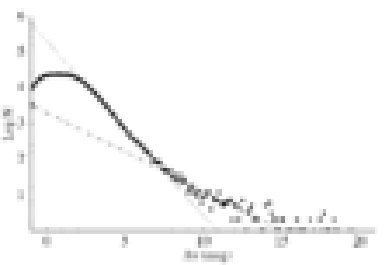}
\caption{\label{app5_1} Best fits (dotted line) to the Column Density
Distribution for the cloud Auriga\,1 (solid line) for a spatial scale of
0.1\,pc. Both fits (for the high and low column density regions) are shown.}
\end{figure}

\begin{figure}
\centering
\includegraphics[width=8.5cm]{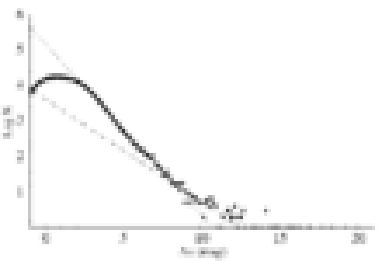}
\caption{As Fig.\,\ref{app5_1} but for the cloud Auriga\,2.}
\end{figure}

\begin{figure}
\centering
\includegraphics[width=8.5cm]{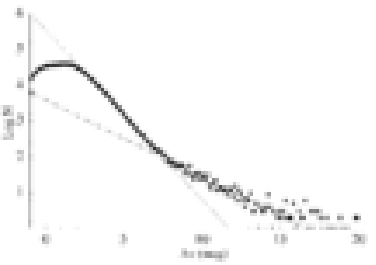}
\caption{As Fig.\,\ref{app5_1} but for the cloud Cepheus.}
\end{figure}

\begin{figure}
\centering
\includegraphics[width=8.5cm]{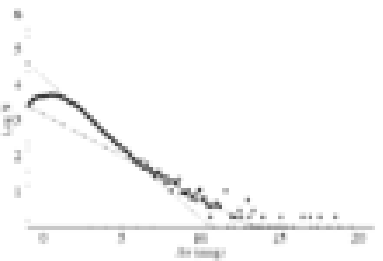}
\caption{As Fig.\,\ref{app5_1} but for the cloud Chamaeleon.}
\end{figure}

\begin{figure}
\centering
\includegraphics[width=8.5cm]{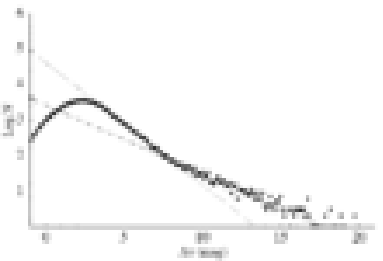}
\caption{As Fig.\,\ref{app5_1} but for the cloud Circinus.}
\end{figure}

\begin{figure}
\centering
\includegraphics[width=8.5cm]{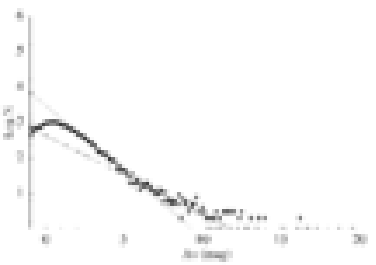}
\caption{As Fig.\,\ref{app5_1} but for the cloud Corona Australis.}
\end{figure}

\begin{figure}
\centering
\includegraphics[width=8.5cm]{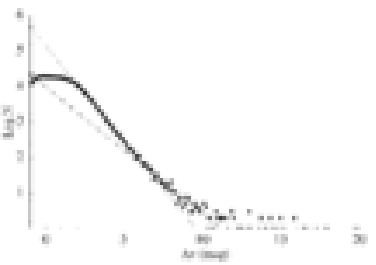}
\caption{As Fig.\,\ref{app5_1} but for the cloud $\lambda$-Ori.}
\end{figure}

\begin{figure}
\centering
\includegraphics[width=8.5cm]{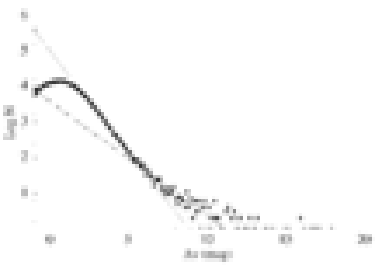}
\caption{As Fig.\,\ref{app5_1} but for the cloud Lupus\,1, 2.}
\end{figure}

\begin{figure}
\centering
\includegraphics[width=8.5cm]{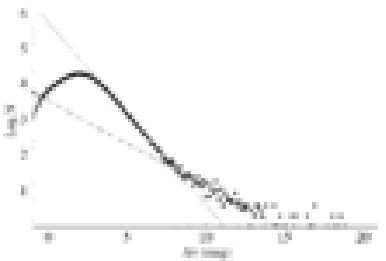}
\caption{As Fig.\,\ref{app5_1} but for the cloud Lupus\,3, 4, 5, 6.}
\end{figure}

\begin{figure}
\centering
\includegraphics[width=8.5cm]{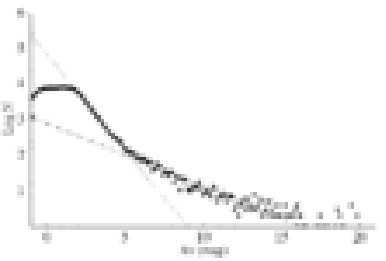}
\caption{As Fig.\,\ref{app5_1} but for the cloud Monoceros.}
\end{figure}

\begin{figure}
\centering
\includegraphics[width=8.5cm]{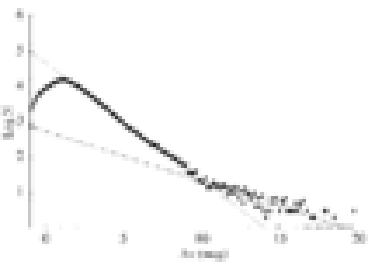}
\caption{As Fig.\,\ref{app5_1} but for the cloud Ophiuchus.}
\end{figure}

\begin{figure}
\centering
\includegraphics[width=8.5cm]{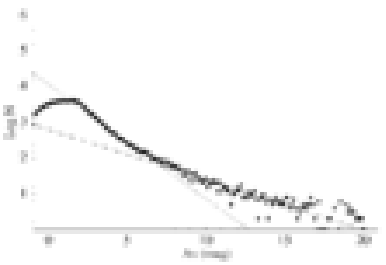}
\caption{As Fig.\,\ref{app5_1} but for the cloud Orion\,A.}
\end{figure}

\begin{figure}
\centering
\includegraphics[width=8.5cm]{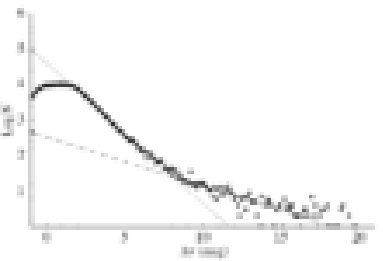}
\caption{As Fig.\,\ref{app5_1} but for the cloud Orion\,B.}
\end{figure}

\begin{figure}
\centering
\includegraphics[width=8.5cm]{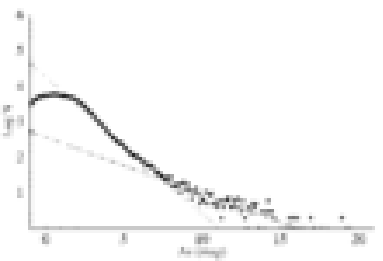}
\caption{As Fig.\,\ref{app5_1} but for the cloud Perseus.}
\end{figure}

\begin{figure}
\centering
\includegraphics[width=8.5cm]{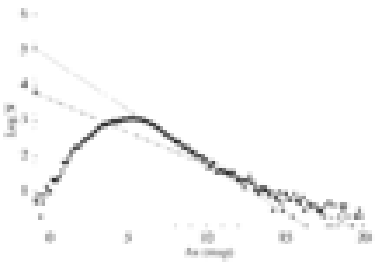}
\caption{As Fig.\,\ref{app5_1} but for the cloud Serpens.}
\end{figure}

\begin{figure}
\centering
\includegraphics[width=8.5cm]{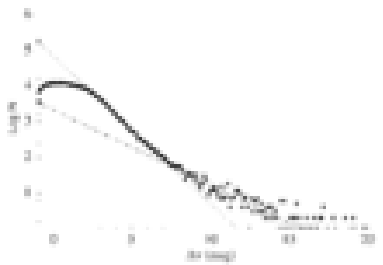}
\caption{As Fig.\,\ref{app5_1} but for the cloud Taurus.}
\end{figure}

\clearpage
\newpage

\subsection{The log(M) vs $A_V$ Mass Distributions}
\label{app6}

\clearpage
\newpage

\begin{figure}
\centering
\includegraphics[width=8.5cm]{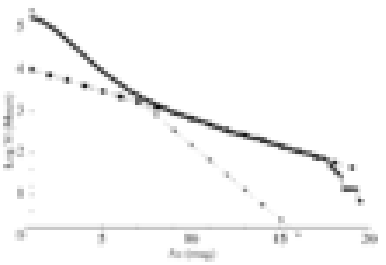}
\caption{\label{app6_1} Best fits (dotted line) to the Mass Distribution for the
cloud Auriga\,1 (solid line) for a spatial scale of 0.1\,pc. Both fits (for the
high and low column density regions) are shown.} 
\end{figure}

\begin{figure}
\centering
\includegraphics[width=8.5cm]{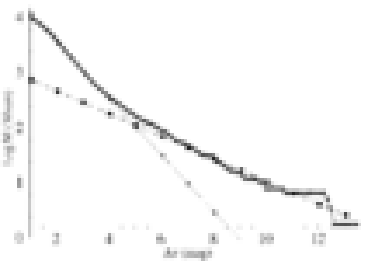}
\caption{As Fig.\,\ref{app6_1} but for the cloud Auriga\,2.}
\end{figure}

\begin{figure}
\centering
\includegraphics[width=8.5cm]{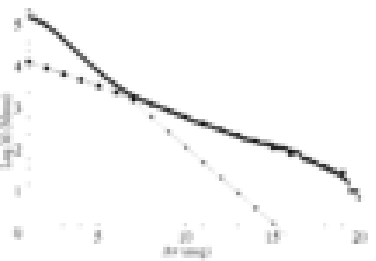}
\caption{As Fig.\,\ref{app6_1} but for the cloud Cepheus.}
\end{figure}

\begin{figure}
\centering
\includegraphics[width=8.5cm]{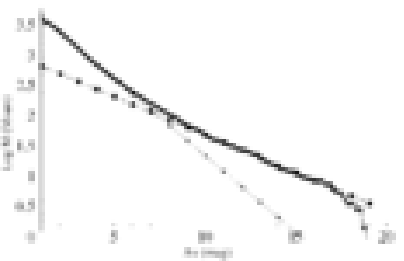}
\caption{As Fig.\,\ref{app6_1} but for the cloud Chamaeleon.}
\end{figure}

\begin{figure}
\centering
\includegraphics[width=8.5cm]{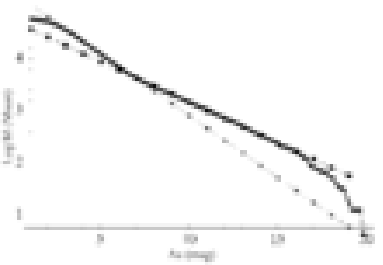}
\caption{As Fig.\,\ref{app6_1} but for the cloud Circinus.}
\end{figure}

\begin{figure}
\centering
\includegraphics[width=8.5cm]{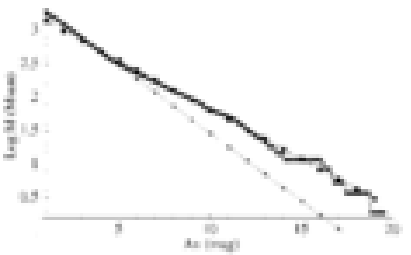}
\caption{As Fig.\,\ref{app6_1} but for the cloud Corona Australis.}
\end{figure}

\begin{figure}
\centering
\includegraphics[width=8.5cm]{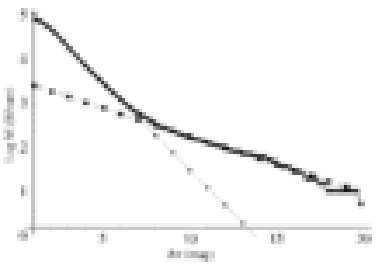}
\caption{As Fig.\,\ref{app6_1} but for the cloud $\lambda$-Ori.}
\end{figure}

\begin{figure}
\centering
\includegraphics[width=8.5cm]{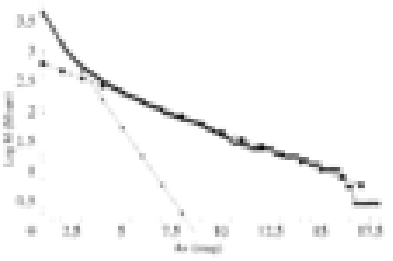}
\caption{As Fig.\,\ref{app6_1} but for the cloud Lupus\,1, 2.}
\end{figure}

\begin{figure}
\centering
\includegraphics[width=8.5cm]{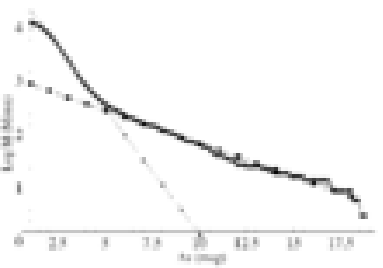}
\caption{As Fig.\,\ref{app6_1} but for the cloud Lupus\,3, 4, 5, 6.}
\end{figure}

\begin{figure}
\centering
\includegraphics[width=8.5cm]{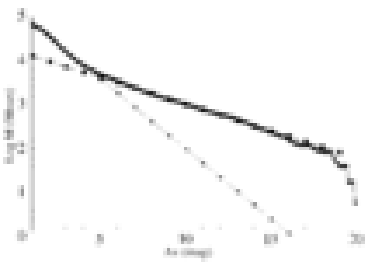}
\caption{As Fig.\,\ref{app6_1} but for the cloud Monoceros.}
\end{figure}

\begin{figure}
\centering
\includegraphics[width=8.5cm]{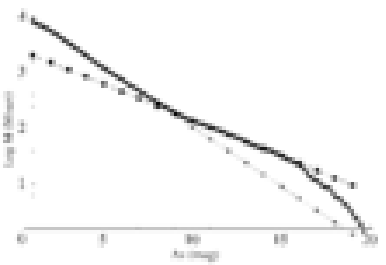}
\caption{As Fig.\,\ref{app6_1} but for the cloud Ophiuchus.}
\end{figure}

\begin{figure}
\centering
\includegraphics[width=8.5cm]{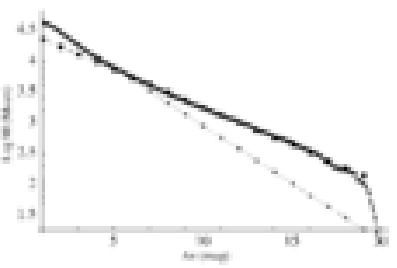}
\caption{As Fig.\,\ref{app6_1} but for the cloud Orion\,A.}
\end{figure}

\begin{figure}
\centering
\includegraphics[width=8.5cm]{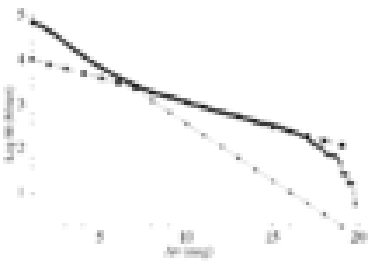}
\caption{As Fig.\,\ref{app6_1} but for the cloud Orion\,B.}
\end{figure}

\begin{figure}
\centering
\includegraphics[width=8.5cm]{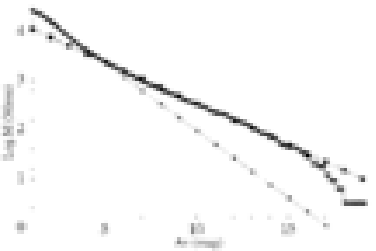}
\caption{As Fig.\,\ref{app6_1} but for the cloud Perseus.}
\end{figure}

\begin{figure}
\centering
\includegraphics[width=8.5cm]{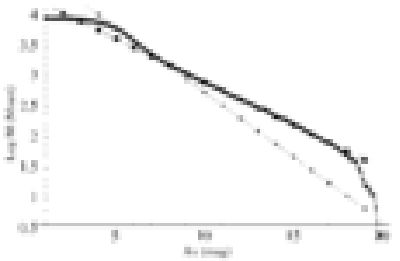}
\caption{As Fig.\,\ref{app6_1} but for the cloud Serpens.}
\end{figure}

\begin{figure}
\centering
\includegraphics[width=8.5cm]{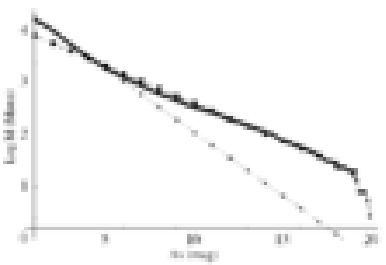}
\caption{As Fig.\,\ref{app6_1} but for the cloud Taurus.}
\end{figure}

\clearpage
\newpage

\end{appendix}

\end{document}